# Multidimensional entropy landscape of quantum criticality

K. Grube*[1], S. Zaum[1], O. Stockert[2], Q. Si[3], H. v. Löhneysen[1,4]

[1] Institut für Festkörperphysik, Karlsruher Institut für Technologie, D-76021 Karlsruhe, Germany.

[2] Max-Planck-Institut für Chemische Physik fester Stoffe, D-01187 Dresden, Germany.

[3] Department of Physics and Astronomy, Rice University, Houston, TX 77005, USA.

[4] Physikalisches Institut, Karlsruher Institut für Technologie, D-76049 Karlsruhe, Germany.

* Correspondence to: kai.grube@kit.edu.

**The Third Law of Thermodynamics states that the entropy of any system in equilibrium has to vanish at absolute zero temperature. At nonzero temperatures, on the other hand, matter is expected to accumulate entropy near a quantum critical point (QCP), where it undergoes a continuous transition from one ground state to another[1,2]. Here, we determine, based on general thermodynamic principles, the spatial-dimensional profile of the entropy $S$ near a QCP and its steepest descent in the corresponding multidimensional stress space. We demonstrate this approach for the canonical quantum critical compound $CeCu_{6-x}Au_x$ near its onset of antiferromagnetic order[2]. We are able to link the directional stress dependence of $S$ to the previously determined geometry of quantum critical fluctuations[3]. Our demonstration of the multidimensional entropy landscape provides the foundation to understand how quantum criticality nucleates novel phases such as high-temperature superconductivity.**

Quantum criticality arises near a second-order phase transition that is driven to zero temperature by competing interactions. For metallic systems, it provides a mechanism to generate new types of electron-derived excitations that are distinct from Landau's Fermi liquid[2]. Because quantum fluctuations are enhanced when the dimensionality is reduced, quantum critical points (QCPs) often arise in anisotropic systems. The quantum critical fluctuations lead to unconventional scaling behavior and the accumulation of entropy at very low $T$, thereby allowing unusual electronic excitations and new phases. The enhanced entropy $S$ upon approaching a QCP has been probed by measurements of the specific heat, and its dependence on pressure was studied by volume thermal expansion. The entropy landscape has been studied up to now using a single tuning parameter[4]. In order to understand how entropy evolves as the system traverses near a QCP, exploration of its profile in a multidimensional parameter space is needed.

Heavy-fermion systems represent prototype settings for QCPs induced by pressure. The latter tunes the hybridization of the almost localized 4f states with the conduction band, thereby tilting the balance in the competition between Ruderman-Kittel-Kasuya-Yosida (RKKY) and Kondo interactions. Previous experiments on quantum critical heavy-fermion systems[5,6] focused on the volume expansivity $\alpha_V$ and volume Grüneisen ratio $\Gamma_V$. Both $\alpha_V/T$ and $\Gamma_V$ were found to diverge as $T \to 0$, indicating a diverging pressure dependence of $S$ and a vanishing energy scale near the QCP as predicted[7]. Spatial anisotropy, a hallmark of many heavy-fermion systems, allows a QCP to be accessed with multiple tuning parameters.

Here we show that, for anisotropic systems, the directional dependence of the thermal expansivity provides a means to determine the spatial-dimensional profile of the thermodynamic singularities near a QCP. We establish a procedure to identify the combination of stresses that aims directly at the QCP and accomplishes the steepest change of the entropy $S$. We thereby can find the optimal way to approach the QCP and *in principle* directly link it with the geometry of the underlying quantum critical fluctuations.

We now specify the quantities of interest to our study. While the specific heat $C$ reveals the $T$ dependence of the entropy $S$, the linear thermal-expansion coefficients are related to its uniaxial pressure dependence: $\alpha_i = \partial \varepsilon_i / \partial T = -V^{-1} \partial S / \partial \sigma_i$. Here, $\varepsilon_i$ and $\sigma_i$ are strain and stress, respectively, along the principal crystallographic axes, for orthorhombic crystal structures: $i = a,b,c$ (see Supplementary Note). $V$ is the molar volume. If a system is dominated by a single energy scale $E^*$ as in a Fermi liquid, $C$ and $\alpha_i$ are proportional to each other. In this case the proportional factor, the Grüneisen ratio, related to the normalized stress dependence of $E^*$, $\Gamma_i(\sigma_i) = V\alpha_i / C = d \ln E^* / d\sigma_i$ is constant. The three components $\alpha_i$ are then proportional to each other, i.e., their anisotropy is temperature independent, as $\Gamma_i / \Gamma_j = \alpha_i / \alpha_j$ with $i,j = a,b,c$. The quantities define the volume expansivity $\alpha_V = \sum_i \alpha_i$ and volume Grüneisen ratio $\Gamma_V = \sum_i \Gamma_i$.

For our study of anisotropic quantum criticality, we choose the heavy-fermion compound $CeCu_{6-x}Au_x$ which is characterized by a strongly anisotropic structure with orthorhombic symmetry, space group *Pmna* (neglecting a minute monoclinic distortion, see Methods), and Ising-like magnetic anisotropy[8]. Consequently, all directional properties exhibit a considerable dependence on the crystal orientation. In addition, inelastic neutron scattering experiments[3] give evidence that the critical, incommensurate magnetic fluctuations at the QCP ($x_c \approx 0.1$) are of quasi two-dimensional (2D) nature and form two sets of correlated planes that are spanned by $[0,1,0]$ and approximately $[0.73, 0, \pm 0.68]$. $CeCu_{5.9}Au_{0.1}$ constitutes therefore an ideal platform to investigate the effect of anisotropic uniaxial pressures on quantum critical behavior.

Figure 1a shows the 4f-electron contribution to the thermal-expansion coefficients $\alpha_i$ of $CeCu_{5.9}Au_{0.1}$ as a function of temperature in the range up to the Kondo temperature, $T_K \approx 6$ K.

All $\alpha_i$ display characteristic features with slightly different positions in temperature. While $\alpha_c$ has a broad maximum at ~ 1.5 K, $\alpha_a$ and $\alpha_b$ show clear shoulders at ~ 0.3 K. In this $T$ range $\alpha_a(T)$ is negative while $\alpha_c(T)$ and $\alpha_b(T)$ are positive. The low-$T$ data below ~ 1 K cover the quantum critical regime. The data above 6 K change their anisotropy reflecting the effect of the known crystalline-electric-field (CEF) excitations, i.e., at $\approx$ 7 meV as found in inelastic neutron scattering and specific-heat measurements[9].

In the quantum-critical temperature regime, the linear thermal expansivities divided by temperature, $\alpha_i/T$ ($i = a,b,c$), are shown along each direction in Fig. 1**b** on a logarithmic $T$ scale. Each component $\alpha_i/T$ tends to diverge towards low temperatures indicating non-Fermi-liquid behavior. The same is also true for the volume thermal-expansion coefficient divided by temperature plotted in Fig. 1**c**. This should be contrasted to a Fermi liquid, where $\alpha_i/T$ is equal to $-V^{-1}d\gamma/d\sigma_i$ with the Sommerfeld coefficient $\gamma = C/T$ expected to be constant at low $T$. The divergences of $\alpha_i/T$ vs. $T$ for $T \to 0$ complement the well-known[10] logarithmic divergence of $\gamma$ vs. $T$ observed in the same compound, also shown in Fig. 1c. However, the divergences of $\alpha_i/T$ are much stronger than that of $\gamma$ as predicted[7]. Indeed, their ratio, the Grüneisen ratios $\Gamma_i = V\alpha_i/C$, diverge for each direction $i = a,b,c$, as demonstrated in Fig. 2a. The same holds for the hydrostatic-pressure Grüneisen ratio $\Gamma_V$ see Fig. 2b.

We now turn to the analysis of the low-$T$ divergences. Because the Sommerfeld coefficient is best described by a logarithmic temperature dependence, we have fitted the linear thermal coefficients in terms of powers of logarithm. For all three directions, we find that the best representation of the data is given by $\alpha_i/T = a_i[\ln^2(T_{0i}/T)]$ with $a_i = -1.01(3)$, 1.92(3), 1.64(3)$\cdot 10^{-6}$ K$^{-2}$, and $T_{0i} = 2.41(14)$, 2.43(7), 8.18(49) K ($i = a,b,c$) (cf. solid lines in Fig. 1b). To obtain the best $\ln^2 T$ fit, a constant (Fermi-liquid like) contribution $a_{0i}/T = -0.60(8)$, 1.67(8), and 1.68(25)$\cdot 10^{-6}$ K$^{-2}$ was assumed. The $T_{0i}$ are somewhat different from but of similar magnitude as $T_0' = 6.33(7)$ K extracted from $\gamma = a'\ln(T_0'/T)$ (ref. 10). $T_0'$ and $T_{0i}$ constitute cut-off temperatures for the quantum critical behavior arising from the competition between Kondo and RKKY interactions and are, therefore, constrained to the Kondo energy scale $T_K$. Thus, the different $T_{0i}$ might signal an anisotropy of the stress dependence of $T_K$ due to the shape of the 4$f$ ground-state wave function. As $T_0'$ and $T_{0i}$ mark the upper limit of the non-Fermi-liquid regime, summing over $\ln^2(T_{0i}/T)$ with different $T_{0i}$ will at $T \ll T_{0i}$ still result in a $\ln^2 T$ dependence of $\alpha_V/T$ as indeed shown by the fit (solid line) in Fig. 1c. The fit parameters for $\alpha_V/T$ are $b = 2.70(5)\cdot 10^{-6}$ K$^{-2}$ and $T_0 = 4.69(22)$ K. Previous work on polycrystalline CeCu$_{5.8}$Ag$_{0.2}$, which shows quantum critical behavior in the specific heat closely resembling that of CeCu$_{5.9}$Au$_{0.1}$, reported a linear dependence of $\alpha_V/T$ vs. $\ln T$ for $0.07 \leq T \leq 0.5$ K (ref. 6).

The temperature dependence of our Grüneisen ratios $\Gamma_i$ (Fig. 2), in line with the expectation, shows an approximate form of $\Gamma_i = a_{\Gamma,i}\ln(T_{\Gamma,i}/T)$. Because of $T_{0i} \neq T_0'$, this dependence is only approximately observed at $T < 1\,\text{K}$. The deviation at higher $T$ is taken into account by adding a corrective term $2c_i + c_1^2 / [\ln(T_0 / T)]$ with $c_i = \ln(T_0' / T_{0i})$.

To shed light on the origin of the logarithmic temperature dependence of $\Gamma_i$, we first note that a divergent Grüneisen ratio is expected for any QCP: scaling dictates its $T$ dependence as $1/T^y$, where the exponent $y$ is the scaling dimension of the operator that tunes through the zero-temperature transition[7]. For a spin-density-wave QCP, the expected exponent is $y = 1$, which is inconsistent with our measured Grüneisen ratio. In local QCP involving a Kondo destruction[11], as arising in an Ising-anisotropic Kondo lattice appropriate for CeCu$_{6-x}$Au$_x$, the corresponding scaling dimension is $y = 0^+$ (see Methods), which is consistent with the observed logarithmic divergence of the Grüneisen ratio. We note that the model of critical spin fluctuations bootstrapped by energy fluctuations proposed $C/T \sim T^{-1/8}$ for the specific heat of a 2D QCP[12,13]: since weak power-law and logarithmic $T$ dependencies are difficult to distinguish, a small-exponent power law is compatible with the $C/T$ data of CeCu$_{5.9}$Au$_{0.1}$; however, this same model predicted $T^{-1}$ dependence for $\alpha_i/T$, which is much stronger than the observed $\ln^2 T$ dependence.

Regardless of the (model-dependent) analytical form of the divergence of the volume Grüneisen ratio, the natural question arises of how an anisotropic system such as CeCu$_{5.9}$Au$_{0.1}$ would respond to stress in arbitrary direction. The above determination of $\alpha_i$ allows us to describe geometrically the components of the entropy derivatives with respect to stress $\partial S / \partial \sigma_i$ (Supplementary Note). For orthorhombic systems, the three normal stresses $\sigma_i$ are linearly independent of each other and span a Cartesian coordinate system. The steepest change of $S$ is given by the gradient that is formed by the components along each axis:

$$\vec{\nabla} S = (\partial S / \partial \sigma_a, \partial S / \partial \sigma_b, \partial S / \partial \sigma_c).$$

This vector determines the stress combination that maximizes the entropy variation. The construction leads to an entropy landscape in the parameter space of directional stresses for each temperature. The result of this construction for CeCu$_{5.9}$Au$_{0.1}$ at $T = 1$ K, is shown in Fig. 3a.

For a vanishingly small but nonzero temperature, the red arrow $\vec{\nabla} S$ describes the steepest slope in the entropy landscape as the system is tuned towards the QCP, as illustrated for a two-dimensional stress space $(\sigma_x, \sigma_y)$ in Fig. 3b. On the other hand, tuning along any direction perpendicular to $\vec{\nabla} S$, marked by the red circle in Fig. 3a, will leave the distance to the QCP as a function of stress unaltered and thus result in a vanishing critical contribution to the thermal expansion along this particular direction. From Fig. 1b we infer from the roughly constant ratios $\alpha_i : \alpha_j$ ($i, j = a, b, c$) that the direction $\vec{\nabla} S$ is only weakly dependent on temperature. We note that exactly at the QCP, $\partial S / \partial \delta$ is strictly zero for any trajectory passing through the QCP at $\delta_c$. At $\delta_c$, $\partial S/\partial \delta$ changes its sign.[14] Thus, our data indicate that the critical concentration $x_c$ is in fact a little larger than $x = 0.1$ for our sample.

The dependence of $S$ on any arbitrary stress combination $\vec{\sigma}_x$ can be expressed as the projection of $\vec{\nabla}S$ onto the unit vector in $\vec{\sigma}_x$ direction:

$$\partial S / \partial \sigma_x = \vec{\nabla}S \cdot \vec{\sigma}_x / \sigma_x.$$

This allows discussing the isotropic and anisotropic contributions to $\vec{\nabla}S$, corresponding to the responses to the hydrostatic pressure and the so-called pure shear stresses, respectively (see Supplementary Note). The hydrostatic pressure is defined as $\vec{p} = p \cdot (1,1,1)$ resulting in the well-known volume thermal expansion: $\alpha_V = -V^{-1}\vec{\nabla}S \cdot \vec{p} / p = -V^{-1} \operatorname{div} S = V^{-1}\sum_i \alpha_i$.

The pure shear stresses are planar stresses, represented by combinations of two perpendicular uniaxial pressures of opposite sign, e.g., $\vec{\sigma}_{ca} = p \cdot (-1,0,1)$. As they are orthogonal to the hydrostatic pressure, $\vec{\sigma}_{(ij)} \cdot \vec{p} = 0$, they only affect anisotropic stress dependences. For an elastically homogeneous solid, their application results in the cancellation of the Poisson effect so that the shape of the solid is changed only in the $(ij)$ plane while the volume remains constant (see Supplementary Note).

This leads us to analyzing the responses to the pure shear stresses for an anisotropic system such as CeCu$_{5.9}$Au$_{0.1}$, which are not accessible by the hydrostatic-pressure or volume-dependent $\alpha_V$ and $\Gamma_V$. These stress combinations are proportional to the differences of the linear thermal-expansion coefficients along different, perpendicular directions $\partial S / \partial \sigma_{(ij)} \propto \alpha_{(ij)} = \alpha_i - \alpha_j$ where the hydrostatic, isotropic contributions $\alpha_V / 3$ cancel each other. Figure 4a shows the resulting differences divided by $T$. They exhibit contrasting scaling behavior. While $\alpha_{(ca)}/T$ and $\alpha_{(ba)}/T$ exhibit divergences with $\ln^2 T$ dependence as do the single components and $\alpha_V/T$, $\alpha_{(cb)}/T$ levels off below 1 K. Correspondingly, $\Gamma_{(cb)} = \Gamma_c - \Gamma_b$ becomes roughly constant at $T < 0.5$ K, thus seemingly pointing to a stability of the Fermi-liquid state for $T \to 0$ although the specific-heat coefficient $\gamma(T)$ clearly demonstrates NFL behavior. This apparent dichotomy can be resolved by noting that none of the stress combinations perpendicular to $\vec{\nabla}S$ (red circle in Fig. 3a) will tune the system to the QCP. In CeCu$_{5.9}$Au$_{0.1}$ it is accidental that one of the pure shear stresses is almost orthogonal to $\vec{\nabla}S$. This is visualized by the fact that $(0,-1,1)$ is close to the direction of the intersection line of the red and blue circles in Fig. 3a.

Knowing the entropy landscape of CeCu$_{5.9}$Au$_{0.1}$, we can ask how the anisotropic stress space connects with the geometry of the anisotropic quantum-critical magnetic fluctuations determined by inelastic neutron scattering[2] (the important differences between stress and strain for anisotropic systems are outlined in the Methods). To address this issue, we project the maximal stress dependence of $S$ onto the pure-shear-stress plane. This is the pure shear stress applied to the $ac$ plane (see Fig. 3a), as can indeed be seen from Fig. 4b showing $\Gamma_{(ca)}$ to be the largest $\Gamma_{(ij)}$. The application of this pure shear stress $\sigma_{(ca)}$ leads to a distortion of the $ca$ plane which is always normal to the planes of quantum-critical fluctuations. This is, in fact, the only stress combination which does not alter the distance between nearest-neighbor Ce atoms, but results in a tilt of the almost orthogonal fluctuating planes against each other (Supplementary Fig. S3). We note that a change of $S$ is also induced by $\sigma_{(ba)}$. Thus, a linear combination of of $\sigma_{(ca)}$ and $\sigma_{(ba)}$ would also be perpendicular to $p$. The components of this vector in the pure-shear-stress plane would be $a_1\,\sigma_{(ca)} + a_2\,\sigma_{(ba)}$ with $a_1/a_2 < 10\%$ as can be seen by the thinner red arrows in Fig. 3a. Physically, $\sigma_{(ba)}$ corresponds to a distortion of the $ab$ plane (with $c = $ const) and thus to a (smaller) change of the inclination angle between fluctuating planes. This merely reflects the fact that $\sigma_{(ca)}$ and $\sigma_{(ba)}$ are not linearly independent. How these observations relate to the microscopic anisotropies of RKKY vs. Kondo interactions requires a detailed determination of the electronic structure, which remains a challenge for the future.

Our determination of the entropy landscape in $CeCu_{5.9}Au_{0.1}$ explicitly demonstrates how entropy climbs as the system evolves towards its fully-exposed quantum critical point in a multidimensional parameter space. The pronounced entropy enhancement renders quantum critical systems highly susceptible to the development of novel phases, such as unconventional superconductivity. This fundamentally promotes the understanding of superconductivity for heavy-fermion systems[15]. Another prominent example along this line is the cuprates[16], where the entropy as a function of hole doping $n_h$ peaks at the value where the $T_c(n_h)$ dome is maximum[17]. This entropy maximum is found well above $T_c$ and, moreover, the anisotropy of the cuprates due to the quasi-2D electronic structure entails a strongly anisotropic stress dependence[18]. Other pertinent systems with strong correlations include the iron pnictides and chalcogenides[19], and organic charge-transfer salts[20]. In all these systems, novel phases emerge close to instabilities that are characterized by a strongly enhanced entropy in a phase space which is spanned up by multiple control parameters.

References and Notes:
1. Coleman, P & Schofield, A. J. Quantum criticality. *Nature* **433**, 226 (2005).
2. Löhneysen, H. v., Rosch, A., Vojta, M. & Wölfle P. Fermi-liquid instabilities at magnetic quantum phase transitions. *Rev. Mod. Phys.* **79**, 1015 (2007).
3. Stockert, O. *et al.* Two-Dimensional Fluctuations at the Quantum-Critical Point of $CeCu_{6-x}Au_x$. *Phys. Rev. Lett.* **80**, 5627 (1998).
4. Rost, A. W. *et al.* Entropy Landscape of Phase Formation Associated with Quantum Criticality in $Sr_3Ru_2O_7$. *Science* **325**, 1360 (2009).
5. Küchler, R. *et al.* Divergence of the Grüneisen Ratio at Quantum Critical Points in Heavy Fermion Metals. *Phys. Rev. Lett.* **91**, 066405 (2003).
6. Küchler, R. *et al.* Grüneisen Ratio Divergence at the Quantum Critical Point in $CeCu_{6-x}Ag_x$. *Phys. Rev. Lett.* **93**, 096402 (2004).
7. Zhu, l., Garst, M., Rosch, A. & Si, Q. Universally Diverging Grüneisen Parameter and the Magnetocaloric Effect Close to Quantum Critical Points. *Phys. Rev. Lett.* **91**, 066404 (2003).

8. Schlager, H., Schröder, A., Welsch, M. & Löhneysen, H. v. Magnetic ordering in CeCu$_{6-x}$Au$_x$ single crystals: Thermodynamic and transport properties. *J. Low Temp. Phys.* **90**, 181 (1993).
9. Stroka, B. *et al.* Crystal-field excitations in the heavy-fermion alloys CeCu$_{6-x}$Au$_x$ studied by specific heat and inelastic neutron scattering. *Z. Phys. B* **90**, 155 (1993).
10. Löhneysen, H. v. *et al.* Non-Fermi-liquid behavior in a heavy-fermion alloy at a magnetic instability. *Phys. Rev. Lett.* **72**, 3262 (1994).
11. Si, Q., Rabello, S., Ingersent, K. & Smith, J. L. Locally critical quantum phase transitions in strongly correlated metals. *Nature* **413**, 804 (2001).
12. Abrahams, E. & Wölfle, P. Critical quasiparticle theory applied to heavy fermion metals near an antiferromagnetic quantum phase transition. *Proc. Natl. Acad. Sci. U.S.A* **109**, 3238 (2012).
13. Abrahams, E., Schmalian, J. & Wölfle, P. Strong-coupling theory of heavy-fermion criticality. *Phys. Rev. B* **90**, 045105 (2014).
14. Garst, M. and Rosch, J. Sign change of the Grüneisen parameter and magnetocaloric effect near quantum critical points. *Phys. Rev. B* **72**, 205129 (2005).
15. Marthur, N. *et al*. Magnetically mediated superconductivity in heavy fermion compounds. *Nature* **394**, 39 (1998).
16. Badoux, S. *et al*. Change of carrier density at the pseudogap critical point of a cuprate superconductor. *Nature* **531**, 210 (2016).
17. Loram, J. W. *et al*., Journal of Physics and Chemistry of Solids 62, 59-64 (2001).
18. Hardy, F. *et al.*, Phys. Rev. Lett. 105, 167002 (2010).
19. Si, Q., Yu, R. & Abrahams, E. High Temperature Superconductivity in Iron Pnictides and Chalcogenides. *Nat. Rev. Mater*. 1, 16017 (2016).
20. Oike, H., Miyagawa, K., Taniguchi, H. & Kanoda, K. Pressure-Induced Mott Transition in an Organic Superconductor with a Finite Doping Level. *Phys Rev. Lett*. 114, 067002 (2015).

**Acknowledgments** We thank C. Meingast, P. Wölfle and L. Zhu for valuable discussions and S. Drobnik for experimental help. The work at Karlsruhe was supported by the DFG Research Unit FOR960 "Quantum Phase Transitions", work at Rice University by the ARO Grant No. W911NF-14-1-0525 and the Robert A. Welch Foundation Grant No. C-1411, with travel support provided by NSF Grant No. DMR-1611392. One of us (Q.S.) graciously acknowledges the support of the Alexander von Humboldt Foundation and the hospitality of the Karlsruhe Institute of Technology. This article was completed during stays of Q. Si and H. v. Löhneysen at the Kavli Institute for Theoretical Physics at University of California, Santa Barbara supported by the NSF Grant No. PHY-1066293, and the Aspen Center for Physics supported by the NSF grant No. PHY-1066293.

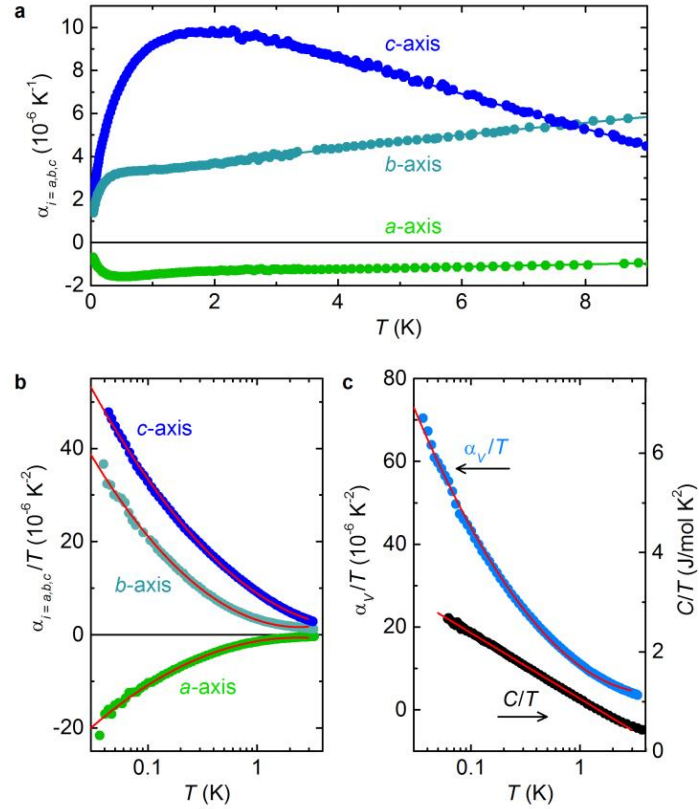

**Figure 1 | Thermal expansion of CeCu$_{5.9}$Au$_{0.1}$. a**, Linear thermal-expansion coefficients $\alpha_i$ as a function of temperature for all three directions $i = a, b, c$ in the orthorhombic notation (see Methods). With decreasing temperature, the varying anisotropy of $\alpha_i$ signals a change of the underlying dominating interactions. **b**, The linear thermal-expansion coefficients $\alpha_i$ from **a** divided by $T$ as a function of ln $T$. **c**, The volume thermal-expansion coefficient divided by $T$ (left-hand scale) and the specific-heat coefficient $C/T$ (right-hand scale) vs. ln $T$. The solid lines are fits to the data as described in the text.

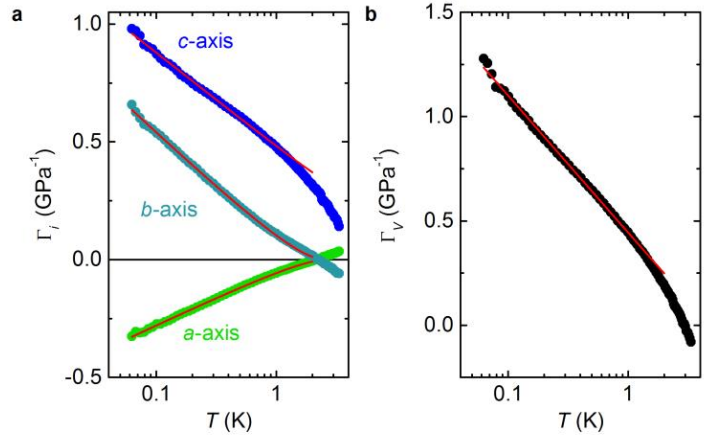

**Figure 2 | Grüneisen ratios of CeCu$_{5.9}$Au$_{0.1}$. a**, The stress Grüneisen ratios $\Gamma_i$ along the three crystal axes as a function of $T$. **b**, The hydrostatic pressure Grüneisen ratio vs. $T$. The solid lines are fits as described in the text.

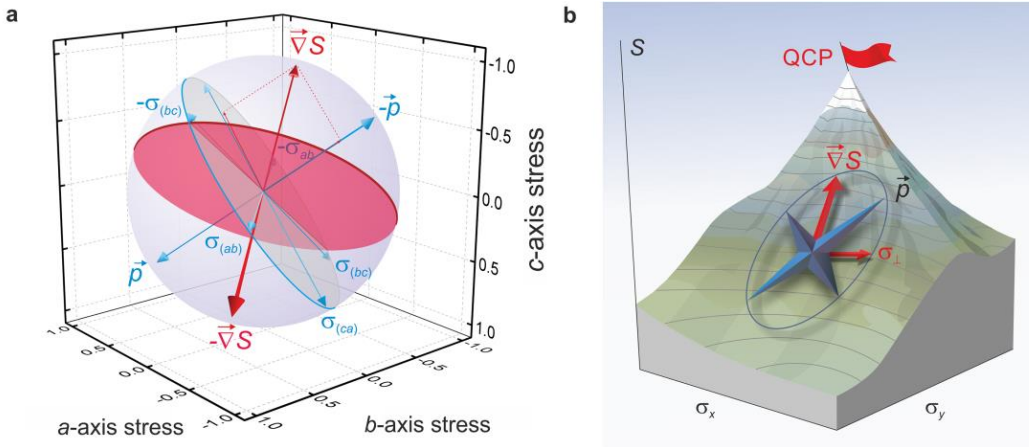

**Figure 3 | Entropy landscape of CeCu$_{5.9}$Au$_{0.1}$ at $T \approx 1$ K. a**, Pictorial illustration of the anisotropic stress dependence of $S$. The red arrow represents the stress combination leading to the steepest slope $\vec{\nabla} S$ aiming at the QCP. All perpendicular stress directions lie within the red plane and $\partial S / \partial \sigma$ vanishes for these directions. The blue arrow indicates the hydrostatic pressure $\vec{p}$. The blue plane perpendicular to $\vec{p}$ represents the possible pure-shear-stress combinations (thinner blue arrows for an orthorhombic Cartesian system $\sigma_{(ij)}$). As shown by the dashed red lines, $\vec{\nabla} S$ has a hydrostatic component and one which is nearly parallel to $\sigma_{(ca)}$. The pure shear stress $-\sigma_{(cb)}$ is close to the intersection line between the two planes and therefore approximately perpendicular to $\vec{\nabla} S$. Accordingly, $\Gamma_{(cb)}$ does not diverge at low $T$ as shown in Fig. 4. **b**, Two-dimensional analog of the entropy landscape discussed for a given temperature. Tuning the system along directions orthogonal to the steepest slope $\vec{\nabla} S$, i.e., following the contour lines, will leave the distance to the QCP unchanged.

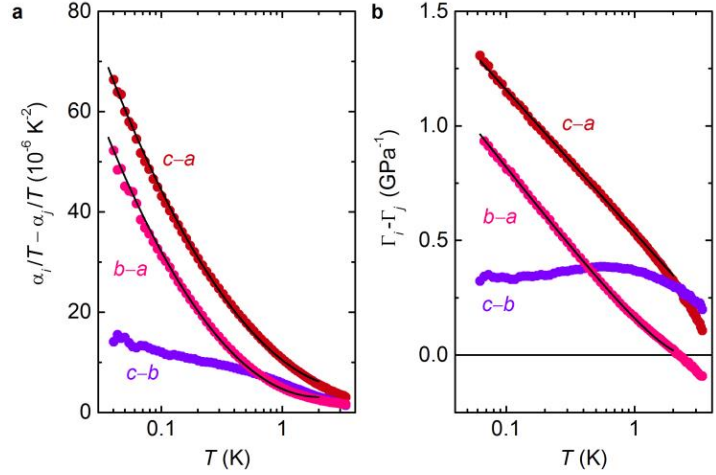

**Figure 4 | Pure shear stress dependences of CeCu$_{5.9}$Au$_{0.1}$ at $T = 1$ K. a**, $\alpha_{(ij)}/T = \alpha_i/T - \alpha_j/T$ with $i \neq j = a,b,c$ as a function of $T$. **b**, The corresponding differences of the Grüneisen ratios $\Gamma_{(ij)} = \Gamma_i - \Gamma_j$ vs. $T$. The solid lines are fits as described in the text.

## Methods:

**Experimental setup.** The thermal-expansion measurements were carried out in a temperature range from 30 mK to room temperature by using a home-built capacitive dilatometer mounted in a dilution refrigerator (30 mK – 9 K) and in a helium gas-flow cryostat (5 – 300 K). The raw thermal-expansion data were corrected for the expansion of the dilatometer by using measurements of Si and Cu single crystals. The investigated CeCu$_{5.9}$Au$_{0.1}$ single crystal was grown by the Czochralski method under high-purity Argon atmosphere, oriented with Laue X-ray diffraction, and spark-cut to yield a cube of approximately 3 mm edge length. For this length, the low-temperature resolution $\Delta L/L$ of the measurements reached $10^{-10}$. The measurements were performed along the three axes of the orthorhombic crystal structure (with orthorhombic axes notation) neglecting the very small monoclinic distortion setting in below $T_s \approx 64$ K with $\varphi_0 = 90.7°$ for $T \ll T_s$ (ref. 21-23). The 4$f$-electron contribution to the thermal expansivity was determined by subtracting the expansivities of an isostructural LaCu$_6$ single crystal[24].

**Pressure vs. volume - stress vs. strain.** Throughout the main part of this paper we have used pressure-dependent thermodynamic quantities, like the pressure-dependent entropy and the Grüneisen ratios $\Gamma_i(\sigma_i)$ as a function of stress. However, historically, the strain (or volume) dependent Grüneisen parameters $\Gamma_i^{(\varepsilon)}(\varepsilon_i) = \sum_{j=1}^{3} c_{ij}\Gamma_j$ have been regarded as more fundamental, because they are not affected by the elasticity of the material—here taken into account by the elastic constants $c_{ij}$—and are related directly to atomic positions or bond lengths. In particular, they are corrected for the Poisson effect, i.e., the dilation (compression) of a solid body perpendicular to the direction of the applied compressive (tensile) stress. For studies of critical phenomena, on the other hand, which have been performed at constant pressure, as our experiments, the use of $\Gamma_i^{(\varepsilon)}$ is inappropriate because here pressure constitutes the control parameter.

In addition, the conversion to $\Gamma_i^{(\varepsilon)}$ demands the knowledge of the elastic constants $c_{ij}$. Their determination requires elaborate measurements, as compounds of lower crystal symmetry are characterized by a large number of independent $c_{ij}$ constants. The elastic behavior of CeCu$_{6-x}$Au$_x$ is described by nine constants, six of which have been published so far for $x = 0$ and partly for 0.1 (refs. 25-28). Between 100 mK and 10 K they change by less than $\Delta c_{ij}/c_{ij} = 10^{-3}$. The elasticity of CeCu$_{6-x}$Au$_x$ is mainly determined by the chemical bonds between the Cu/Au atoms and the outer Ce $6s^2$ and $5d^1$ electrons. Consequently, the $c_{ij}$ can be regarded as temperature independent and corresponding linear combinations of $\Gamma_i^{(\varepsilon)}$ show the same $T$ dependence as $\Gamma_i(\sigma_i)$. It is worth noting that the weak coupling of the critical fluctuations to the crystal lattice ensures that the scaling behavior is not obscured by a softening of the lattice.

The conversion from stress to strain dependences does not qualitatively change the overall behavior of CeCu$_{6-x}$Au$_x$ as can be checked if all $c_{ij}$ are known. We therefore have determined the remaining elastic constants $c_{12}$, $c_{13}$, and $c_{23}$ by employing measurements of the linear compressibilities $k_i = -\partial \varepsilon_i / \partial p$ (with $i = a,b,c$ in Voigt's notation) by using a miniaturized capacitive dilatometer built into a gas pressure cell[29,30]). To ensure hydrostatic pressure conditions, we used helium as pressure transmitting medium. For CeCu$_6$ and $T = 10$ K, the measured linear compressibilities amount to $k_a = 2.884 \times 10^{-3}$ GPa$^{-1}$, $k_b = 4.45 \times 10^{-3}$ GPa$^{-1}$, and $k_c = 3.435 \times 10^{-3}$ GPa$^{-1}$. The elastic constants hardly change between $x = 0$ and 0.1 (ref. 28). To complete our measurements, we have chosen the $c_{ij}$ data from Weber et al.[25] due to their better agreement with ours and the low-temperature measurements from Finsterbusch et al.[28]. By using these constants, we can estimate the dependence of the lattice parameters on arbitrary combinations of stresses. For a pure shear stress of the $ca$ plane $\vec{\sigma}_{(ca)} = (\sigma_{(ca)}/\sqrt{2})(-1,0,1)$, the stress dependence along the *a* axis $\partial \varepsilon_a / \partial \sigma_{(ca)}$ is approximately equal to $-\partial \varepsilon_c / \partial \sigma_{(ca)}$ and with $\approx 0.02$ GPa$^{-1}$ clearly larger than the change of the entire volume strain $\partial \varepsilon_V / \partial \sigma_{(ca)} \approx$ 0.0005 GPa$^{-1}$. Pure shear stresses applied to other planes result in similar, negligibly small changes of $\varepsilon_V$. Therefore, the pure shear stresses can be approximately understood as pure shear strains, i.e., distortions without a sizable volume change.

The influence of the elastic properties on the anisotropy of the linear thermal-expansion coefficients and the Grüneisen ratios can be determined from the related strain dependences of the entropy $\partial S / \partial \varepsilon_i = V \cdot \sum_{j=1}^{3} c_{ij} \alpha_j$ and the aforementioned Grüneisen parameters $\Gamma_i^{(\varepsilon)}$. At low temperature they show the same anisotropy as the stress dependent quantities. In conclusion, both stress and strain dependences can be explained by a combination of a hydrostatic, isotropic volume effect and a high sensitivity to distortions of the *ca* plane.

**Functional form of the Grüneisen ratio divergences from scaling analysis.** We consider the Kondo-lattice model as specified by the Hamiltonian

$$H = \sum_{ij\sigma} t_{ij} c_{i\sigma}^\dagger c_{j\sigma} + \sum_i J_K \mathbf{S}_i \cdot \mathbf{s}_{c,i} + \sum_{ij} \frac{I_{ij}}{2} \mathbf{S}_i \cdot \mathbf{S}_j. \quad (1)$$

Here $\mathbf{S}_i$ and $\mathbf{s}_{c,i}$, respectively, describe a local moment at site $i$ and the spin of the conduction electrons at the same site, and $t_{ij}$ is the tight-binding hopping parameter whose spatial Fourier transform yields the band dispersion $\varepsilon_\mathbf{k}$. The parameters $J_K$ and $I_{ij}$ are the Kondo and RKKY exchange interactions, respectively. Within the extended dynamical mean-field theory[31], the local properties of this lattice model are determined through a self-consistent Bose-Fermi Kondo model:

$$H_{\text{imp}} = J_K \mathbf{S} \cdot \mathbf{s}_c + \sum_{p,\sigma} E_p c_{p\sigma}^\dagger c_{p\sigma} + g \sum_p \mathbf{S} \cdot \left( \vec{\phi}_p + \vec{\phi}_{-p}^\dagger \right) + \sum_p w_p \vec{\phi}_p^\dagger \cdot \vec{\phi}_p, \quad (2)$$

where $E_p$ and $w_p$ are the dispersions determined self-consistently for the fermionic bath ($c_{p\sigma}$) and bosonic bath ($\vec{\phi}_p$), respectively. The locally quantum critical point with a Kondo destruction arises in the self-consistent solution to the model if the magnetic fluctuations are two-dimensional[31]. At the locally quantum critical point, the density of states of the fermionic bath is

$$\sum_p \delta(\omega - E_p) = N_0, \quad (3)$$

i.e., the bare density of states of the conduction electrons, whereas the spectral function of the bosonic bath is

$$\sum_p \left[ \delta(\omega - w_p) - \delta(\omega + w_p) \right] = (K_0^2/\pi) |\omega|^\gamma \, \text{sgn}\,\omega \qquad \text{for } |\omega| < \Lambda, \tag{4}$$

where $\Lambda$ is a cutoff frequency. Self-consistently, the power exponent of $\omega$ is $\gamma = 0^+$.

To determine the scaling dimension of the most relevant operator for the Kondo destruction, we use an $\varepsilon$-expansion procedure discussed earlier[32]. Here $\varepsilon \equiv 1 - \gamma$ which in the end is set to its self-consistent value $1^-$. Unlike the critical exponent for the local spin susceptibility, the exponent pertinent to our analysis here turns out to depend on the symmetry in the spin space. To show this, we allow the exchange interactions in the longitudinal (between the $Z$ components) and transverse (between the $XY$ components) channels to be different. For the Kondo exchange interaction, they are given by $J_K^Z$ and $J_K^\perp$, respectively. Likewise, for the spin-boson coupling, they correspond to $g^Z$ and $g^\perp$, respectively. These are the four coupling constants, $A_i$ $(i = 1, 2, 3, 4)$, that come into the RG analysis.

The beta functions $\beta_{A_i}$ for the anisotropic Bose-Fermi Kondo problem with $xy$ invariance are given in Ref. 32. The fixed points are determined by setting $\beta_{A_i} = 0$. The scaling dimensions can be obtained by solving for the eigenvalues of the $4 \times 4$ matrix $-\partial \beta_{A_i}/\partial A_j$, evaluated at the fixed point. To order $\varepsilon^2$, the largest eigenvalue of the matrix is

$$\lambda = \frac{\sqrt{5}-1}{2} \varepsilon + \frac{5\sqrt{5}-9}{8(5+\sqrt{5})} \varepsilon^2 \approx 0.618034\varepsilon + 0.0376644\varepsilon^2. \tag{5}$$

It is approximately equal to 0.62 and 0.66 to the linear and quadratic orders in $\varepsilon$, respectively. For the $\varepsilon = 1^-$ limit, this corresponds to the exponent $\eta \approx 0.66$ of the temperature dependence of the Grüneisen ratio $\Gamma \propto T^{-\eta}$. This value is consistent with the Grüneisen ratio observed in YbRh$_2$Si$_2$ (ref. 5), in which the magnetic anisotropy has the $xy$ form. For the Ising case appropriate for CeCu$_{6-x}$Au$_x$ (ref. 33), the $\varepsilon$-expansion is set up in terms of RG equations that generalize those for the Kosterlitz-Thouless case (for $\varepsilon = 0$) to nonzero $\varepsilon$. The largest eigenvalue is $\lambda = \sqrt{2\varepsilon}$, to the linear order in $\varepsilon$. For small $\varepsilon$, this result compares well with numerical results[34]. Here, however, the corrections from the higher orders are difficult to determine. For the $\varepsilon = 1^-$ limit, the numerical result[34] is that $\lambda$ approaches $0^+$. This corresponds to $\eta = 0^+$, which is consistent with a logarithmically divergent Grüneisen ratio towards low $T$ as experimentally observed here.

**References**

21. Grube, K., Fietz, W. H., Tutsch, U., Stockert, O. & Löhneysen, H. v. Suppression of the structural phase transition in CeCu$_6$ by pressure and Au doping. *Phys. Rev. B* **60**, 11947 (1999).

22. Schröder, A., Aeppli, G., Bucher, E., Ramazashvili, R. & Coleman, P. Scaling of Magnetic Fluctuations near a Quantum Phase Transition. *Phys. Rev. Lett.* **80**, 5623 (1998).

23. Gratz, E., Bauer, E., Nowotny, H., Mueller, H., Zemirli, S. & Barbara, B. Lattice distortion in CeCu$_6$. *J. Magn. Magn. Mater.* **63 & 64**, 312 (1987).

24. Drobnik, S. Thermische Ausdehnung und Magnetostriktion von CeCu$_{6-x}$Au$_x$ bei sehr tiefen Temperaturen. (Ph. D. thesis, Universität Karlsruhe, 2006).

25. Weber, D. Yoshizawa, M., Kouroudis, I., Lüthi, B. & Walker, E. Electron-Phonon Coupling in the Heavy-Fermion Compound CeCu$_6$. *Europhys. Lett.* **3**, 827 (1987).

26. Suzuki, T., Goto, T., Tamaki, A., Fujimura, T., Ōnuki, Y. & Komatsubara, T. Elastic Soft Mode and Crystalline Field Effect of Kondo Lattice Substance; CeCu$_6$. *J. Phys. Soc. Jpn.* **54**, 2367 (1985).

27. Goto, T., Suzuki T., Fujimura T., Ōnuki, Y. & Komatsubara, T. Elastic properties of the Kondo lattice compound CeCu$_6$. *J. Magn. Magn. Mater.* **63**, 309 (1987).

28. Finsterbusch, D., Willig, H., Wolf, B., Bruls, G., Lüthi, B., Waffenschmidt, M., Stockert, O., Schröder, A. & Löhneysen, H. v. Thermodynamic properties of CeCu$_{6-x}$Au$_x$: Fermi-liquid vs. non-Fermi-liquid behaviour. *Annalen der Physik* **508**, 184 (1996).

29. K. Grube, Thermal expansion of C$_{60}$ single crystals under pressure. (Ph. D. thesis, Universität Karlsruhe 1995, Report FZKA 5611, Forschungszentrum Karlsruhe).

30. Fietz, W. H., Grube, K. & Leibrock, H. Dilatometry under high pressure. *High Pressure Research*, **19**, 373 (2000).

31. Si, Q., Rabello, S., Ingersent, K. & Smith, J. L. Local fluctuations in quantum critical metals. *Phys. Rev. B* **68**, 115103 (2003).

32. Zhu, L. & Si, Q. Critical local-moment fluctuations in the Bose-Fermi Kondo model. *Phys. Rev. B* **66**, 024426 (2002).

33. Tomanic, T., Hamann, A. & Löhneysen H. v. Anisotropy of the magnetic susceptibility of CeCu$_{6-x}$Au$_x$ near the quantum phase transition. *Physica B* **403**, 1323 (2008).

34. Glossop, M. T. & Ingersent, K. Kondo physics and dissipation: A numerical renormalization-group approach to Bose-Fermi Kondo models. *Phys. Rev. B* **75**, 104410 (2007).